\documentclass{llncs}
\usepackage{makeidx}  % allows for indexgeneration
\usepackage{graphicx}
\usepackage{caption}
\usepackage{amsmath} 
\usepackage{makecell}
\usepackage{lineno}
\usepackage{algorithm}
\usepackage{algorithmic}
\usepackage{subfig}
\usepackage{blindtext}
\usepackage{epstopdf}
\usepackage{mathtools}
\usepackage{multirow}
\usepackage{amssymb}
\usepackage{booktabs}
\usepackage{tabularx} 
\usepackage{tabu}

\begin{document}
\frontmatter          % for the preliminaries
\pagestyle{headings}  % switches on printing of running heads
%\addtocmark{Hamiltonian Mechanics} % additional mark in the TOC

\mainmatter              % start of the contributions
\title{Fast Prospective Detection of Contrast Inflow in X-ray Angiograms with Convolutional Neural Network and Recurrent Neural Network}
\titlerunning{Fast Prospective Detection of Contrast Inflow}  % abbreviated title (for running head)
\author{Hua Ma \and Pierre Ambrosini \and Theo van Walsum}
% index{Ma, Hua}
% index{Ambrosini, Pierre}
% index{van Walsum, Theo}
%
\authorrunning{H. Ma et al.} % abbreviated author list (for running head)
\tocauthor{*}
%
%\institute{Biomedical Imaging Group Rotterdam, Departments of Medical Informatics and \\Radiology \& Nuclear Medicine, Erasmus MC, Rotterdam, The Netherlands}
\institute{Biomedical Imaging Group Rotterdam, Erasmus MC, Rotterdam, The Netherlands\\ \email{h.ma@erasmusmc.nl}}

\maketitle              % typeset the title of the contribution

%
% -----Abstract-----
%
\begin{abstract}

% purpose
\noindent Automatic detection of contrast inflow in X-ray angiographic sequences can facilitate image guidance in computer-assisted cardiac interventions. In this paper, we propose two different approaches for prospective contrast inflow detection. The methods were developed and evaluated to detect contrast frames from X-ray sequences. The first approach trains a convolutional neural network (CNN) to distinguish whether a frame has contrast agent or not. The second method extracts contrast features from images with enhanced vessel structures; the contrast frames are then detected based on changes in the feature curve using long short-term memory (LSTM), a recurrent neural network architecture. Our experiments show that both approaches achieve good performance on detection of the beginning contrast frame from X-ray sequences and are more robust than a state-of-the-art method. As the proposed methods work in prospective settings and run fast, they have the potential of being used in clinical practice.

\end{abstract}

%
%\begin{keywords}
%X-ray angiogram, contrast inflow, prospective detection, recurrent neural network, convolutional neural network
%\end{keywords}

%
% -----Introduction-----
%
\section{Introduction}
\label{sec:intro}

% Clinical background and motivation

% interventions, images, contrast
\noindent During percutaneous coronary interventions (PCI), X-ray angiography (XA) is commonly used by clinicians to identify the sites of plaque and navigate devices through the arteries of patients with advanced coronary artery disease. As X-ray imaging has poor soft tissue contrast, coronary arteries are normally visualized by injecting radio-opaque contrast agent in the vessels.

\indent Approaches for improving image guidance in such procedures have been reported, for example fusion of coronary models from CTA~\cite{ruijters2009vesselness-based}. Such methods can only be applied if vessels are visible in the XA, thus automated application of such methods requires detection of presence of contrast agent. Similarly, automated detection of catheter and guidewires, which can also be used for virtual roadmapping~\cite{baka2015respiratory}, is generally only possible in non-contrast enhanced frames. Therefore, an automatic way to detect contrast inflow online is relevant for further automating advanced image guidance methods for coronary interventions, reducing interactions of clinicians with computers during procedures.

% related works
\indent Existing works for detection of contrast inflow in X-ray images fall into two categories: enhancement-based and learning-based. Enhancement-based methods \cite{condurache2004fast,liao2013integrated,liao2013automatic,you2011spatio,zhao2013robust} enhance contrasted structures, followed by a step to extract features that indicate the change of contrast throughout the sequence. The contrast-enhanced frames are then detected via analysis of the feature. Learning-based approaches \cite{chen2011robust,hoffmann2015robust} train a classifier to detect contrast or non-contrast frames based on handcrafted image features. Among these works, \cite{liao2013integrated,you2011spatio,chen2011robust} need an entire sequence to detect contrast inflow, and thus only work retrospectively. \cite{liao2013automatic} does not rely on a complete sequence, but retrospectively runs on a sliding segment of a few new X-ray frames, thus there is a trade-off between the possible delay of the contrast inflow detection and the overall processing efficiency. In addition, this method was designed specifically for TAVI procedures on aorta: their contrast detection method involves aligning a predefined aorta shape model to X-ray images and a step of TEE probe detection, which is not relevant for coronary interventions. \cite{zhao2013robust} uses a heuristic approach to detect the first contrast-enhanced frame from X-ray sequences of left atrium (LA) used for electrophysiology (EP) ablation procedures. \cite{hoffmann2015robust} developed a learning-based framework on X-ray images of LA for EP procedures. The method used a SVM classifer with the heuristic features introduced in \cite{condurache2004fast} and \cite{zhao2013robust}. Out of these methods, \cite{condurache2004fast} is the only one that may be directly used for coronary interventions and work in prospective settings.

% Our contributions
\indent The purpose of our work is to develop and evaluate solutions for prospective detection of contrast inflow in XA images that can fit into the clinical work-flow of coronary interventions. Specifically, we aim at prospectively detecting if a frame has contrast agent. To this end, two different approaches were developed. Due to the exceptional performances that convolutional neural networks (CNN) have in image classifications \cite{lecun2015deep}, and medical applications, such as tissue segmentation and surgical tools detection \cite{greenspan2016guest}, we propose a learning-based method using CNN to classify each frame of an XA sequence into two classes: with or without contrast. Additionally, we propose a hybrid of enhancement- and learning-based. It computes a temporal contrast feature from vessel-enhanced sequences based on which contrasted frames are detected with long short-term memory (LSTM) \cite{hochreiter1997long}, a recurrent neural network (RNN) architecture. To the best of our knowledge, this is the first work that applies deep learning for contrast inflow detection in X-ray images. To validate the detection, the position of the beginning contrast frame (BCF) \cite{chen2011robust,hoffmann2015robust} in a sequence (where contrast starts being visible) was used in the experiment.

%
% -----Method
%
\section{Methods}
\label{sec:methods}

\subsection{The CNN-based method}
\label{subsec:learning-based}
	
\noindent Let $S = \{I_1, I_2, \ldots, I_n\}$ denote a sequence of $n$ frames in which $I_c$ is the beginning contrast frame. All frames $I_1$, \ldots, $I_{c-1}$ are associated with the label ``without contrast". The other frames $I_c$, \ldots, $I_n$ have the label ``with contrast".

\begin{figure}[t]
	\centering
	\includegraphics[width=0.75\textwidth]{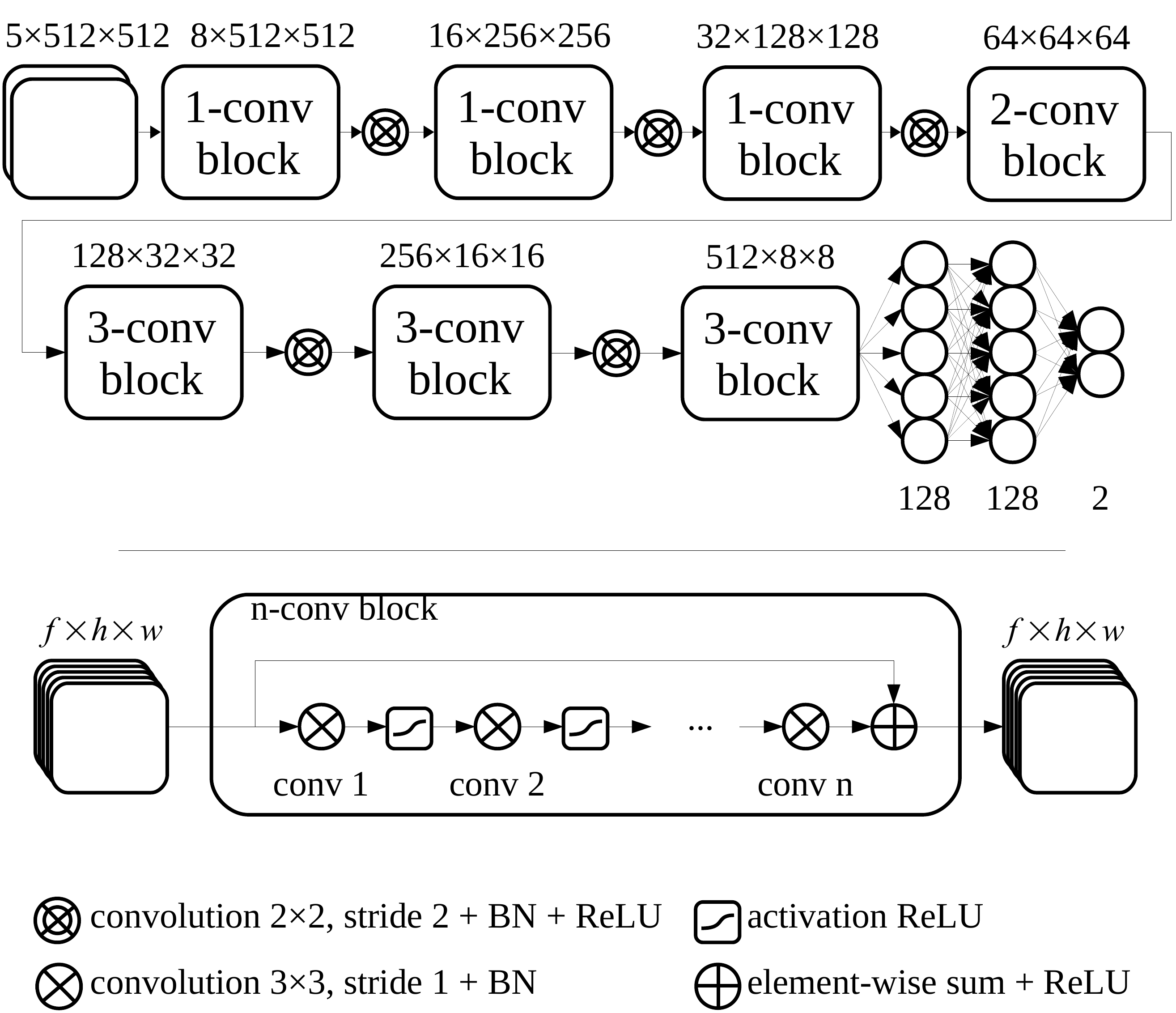}
	\caption{The neural networks (top) connects the 5 input images (the first, the current and its 3 previous X-ray frames) to the 2 output nodes (``with contrast" and ``without contrast"). The model consists of several \textit{n-conv} blocks. They are a succession of CNNs with a skip connection between the input and the output of the block (bottom). $f \times h \times w$ is the dimension of the data (feature number times image height times image width) .}
	\label{fig:cnn_isbi}
\end{figure}

\indent In order to classify the fluoroscopic frames, we used a CNN to learn the difference between the contrast frame and non-contrast frame (Fig.~\ref{fig:cnn_isbi}, top). The input of the CNN has 5 images: the current frame $I_i$ to be classified, its 3 previous frames $I_{i-1}$, $I_{i-2}$, $I_{i-3}$, and the first frame $I_1$ (normally non-contrasted). There are 7 intermediate layers directly after the input layer, each of which has a \textit{n-conv} block with $n$ consecutive convolutions (Fig.~\ref{fig:cnn_isbi}, bottom). The last \textit{n-conv} block is connected with two fully-connected layers. The final output is a softmax layer with two nodes: ``with contrast" and ``without contrast". The model was trained with binary cross-entropy as the loss function. In order for a faster convergence, batch normalization was used after every convolution, residual connection at every layer and the strided convolutions instead of pooling layers. 

\indent To detect the BCF of an XA sequence online using the trained model, frames of the sequence were classified one by one in a chronological order. The first frame labeled as ``with contrast" in the sequence is considered as BCF.

% the neural network figure

\subsection{The RNN-based method}
\label{subsec:enhancement-based}

\begin{figure}[t]
	\centering
	\includegraphics[width=\textwidth]{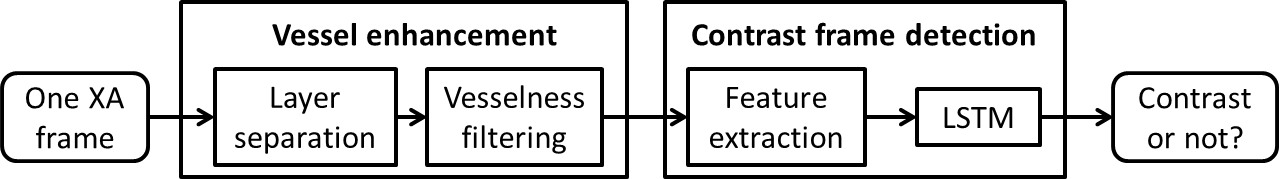}
	\caption{The overview of the RNN-based method.}
	\label{fig:rnn-overview}
\end{figure}

The RNN-based method consists of two major steps: vessel enhancement and contrast frame detection. An overview of this method is illustrated in Fig. \ref{fig:rnn-overview}. \\

%\subsubsection{Vessel enhancement}

\noindent \textbf{Vessel enhancement} The vessel enhancement step is crucial for accurate approximation of contrast changes in XA sequences. This step removes most non-vessel background structures using a previously developed online layer separation technique \cite{ma2016anonymous} followed by multi-scale Frangi-vesselness filtering \cite{frangi1998multiscale}.

% layer separation
\indent The online layer separation method prospectively separated an XA sequence into three layers: a breathing layer, a quasi-static background layer, and a vessel layer in which vessels have better visibility. First, the breathing layer was separated via morphological closing. After this layer was removed from the original image, online robust PCA (OR-PCA) \cite{feng2013online} was applied to separate the low-rank quasi-static layer and sparse vessel layer through alternatively projecting the new data sample (frame) to the underlying low-rank subspace basis and updating the basis using the new estimation of the layers. After layer separation, the structures that may cause artefacts in the next step, such as diaphragm, spine, were removed from the vessel layer.

% Frangi filtering
\indent Following the layer separation, a multi-scale vesselness filter \cite{frangi1998multiscale} was applied on the separated vessel layer to further enhance the tubular structures. In the end, after the vessel enhancement step, for each incoming frame, a new image was created where vessel structures are enhanced. \\

%\subsubsection{Contrast frame detection}

\noindent \textbf{Contrast frame detection} Once the image with enhanced vessel structures is obtained, the feature that indicates the level of contrast agent was extracted from the image. In this work, we used the average pixel intensity of the complete vessel-enhanced image as the contrast feature. This results in a 1D signal for a complete sequence.

\indent The last step is to detect contrast frames from the previously obtained 1D contrast signal. In order to fully use the temporal relation between frames, each signal point is classified as ``contrast" or ``non-contrast" with a recurrent neural network. The long short-term memory (LSTM) network \cite{hochreiter1997long} was used due to its good performance on modeling long-term temporal relations in time-series data. 

\begin{figure}[t]
	\centering
	\includegraphics[width=0.8\textwidth]{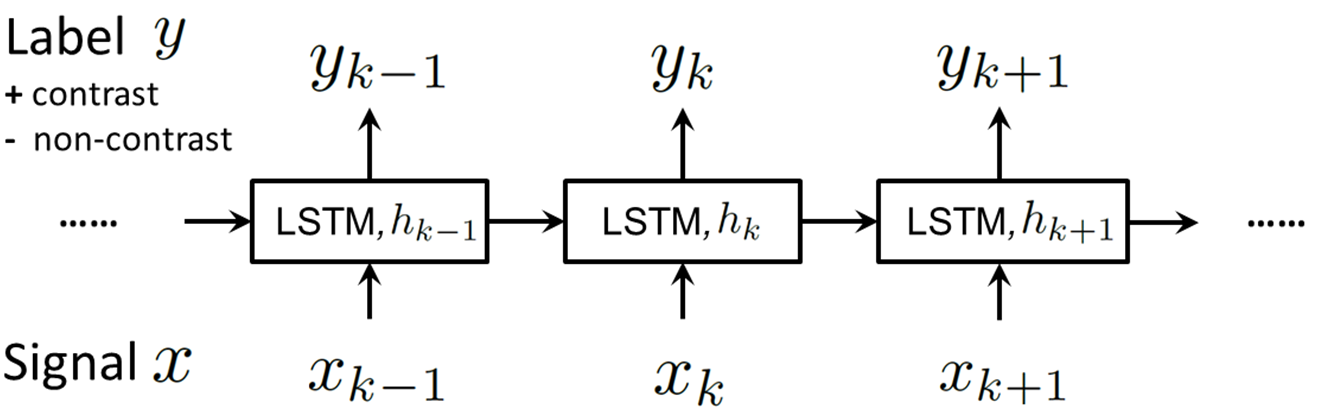}
	\caption{Each signal point is classified from a contrast frame or a non-contrast frame with a LSTM network.}
	\label{fig:lstm}
\end{figure}

\indent Let $x_k$ denote the feature for the \textit{k}th frame $I_k$. The single-direction LSTM takes $x_k$ as the input. A hidden state $h_k$ in the LSTM network is recurrently updated through nonlinear interactions between the input signal $x_k$, the LSTM units and its state of the last time point $h_{k-1}$. The output label $y_k$ of $x_k$ is the outcome of a nonlinear function of $h_k$. This process is illustrated in Fig. \ref{fig:lstm}.

%
% -----Experiments
%
\section{Experiments}
\label{sec:experiments}

We retrospectively obtained anonymized data that was acquired during clinical routine with a Siemens AXIOM-Artis biplane system. The data were 120 XA sequences from 26 patients who underwent a PCI procedure. The frame rate of all sequences is 15 frames per second. The length of sequence varies from 24 to 244 frames. The size of images in our dataset are $512\times512$, $600\times600$, $776\times776$ and $1024\times1024$. In all sequences, contrast inflow can be observed. In our experiments, 40 sequences from 20 patients were used as training data, the 80 sequences from the other six patients were used for validation.

% learning-based method parameter setting
\indent For the CNN-based method, all images were resized to $512\times512$ before training. The parameters of the CNN model were optimized using stochastic gradient descent with a learning rate 0.0001, a decay of 0.0005 and a momentum of 0.99. The model was trained with a batch size of 15 during 33,000 iterations. For each sequence, the six frames before and after the BCF were chosen to ensure an even number of contrast and non-contrast training images. The BCF was discarded to assist the CNN to learn more differences between contrast and non-constrat frames. As the dataset used to train the model is small, data augmentation was applied during the training to virtually create more data: translation (+/- 100 pixels), rotation (+/- 5 degrees), scaling (+/- factor 0.1), intensity shift (+/- 0.2), Gaussian noise ($\sigma_g =$ 0.01) on the normalized image between 0 and 1, and vertical flip were used to transform images. 

% enhancement-based approach parameter setting
\indent For the RNN-based method, we manually tuned the parameters based on visual check and quantitative evaluation on the training data; the same parameters were used for testing. The images were first down-scaled 2 or 4 times to $256\times256$ or $300\times300$ or $388\times388$ depending on the original image size for speeding up the image processing. The parameters for layer separation were set following the approach with the sliding window option in \cite{ma2016anonymous} using the closed-form solution of OR-PCA. To improve the convergence of OR-PCA, we used a mini-batch of 5 frames (before contrast agent was injected) to get an initial estimate of the low-rank subspace basis. This was done using the layer separation method in \cite{ma2015layer} with fast principal component pursuit \cite{rodriguez2013fast}. The scale of Frangi vesselness filter was set ranging from 0.6 $mm$ to 2.8 $mm$ according to the size of coronary arteries. The $\beta$ and $c$ parameter of the vesselness filter were 0.5 and 15. The dimension of LSTM units was set to 7 with a dropout probability being 0.2. The nonlinear activation function of the hidden layer is sigmoid function. The LSTM network was trained using RMSprop optimizer with a learning rate being 0.005 during 100 epochs. At last, the BCF was detected as the first frame in a sequence being classified as contrasted by LSTM.

% state of the art: Condurache et al. 2004, but not Hoffmann et al. 2015
\indent In the experiments, we also compared our methods with the state-of-the-art approach of Condurache et al.\cite{condurache2004fast}. For setting the parameters of the method, the first 3 feature values from non-contrast frames were modeled as a Gaussian $N_0(\mu_0,\sigma_0^2)$. The threshold $T$ for choosing contrast frames was set to $\mu_0 + 3 \sigma_0$.

% evaluation
\indent The evaluation metric we used is the absolute difference between the frame index of the ground truth BCF and the frame predicted by different methods. 

\indent The image processing steps in the RNN-based method and the method of Condurache et al. were implemented in MATLAB with a single CPU core (Intel Core i7-4800MQ 2.70 GHz). LSTM and CNN were implemented in Keras with TensorFlow as backend. LSTM was running on the CPU due to its small dimension. CNN was trained and tested on an Nvidia GeForce GTX 1080 GPU.

%
% -----Results
%
\section{Results and discussion}
\label{sec:results-discussion}

\begin{figure}[t]
	\centering
	\includegraphics[width=\textwidth]{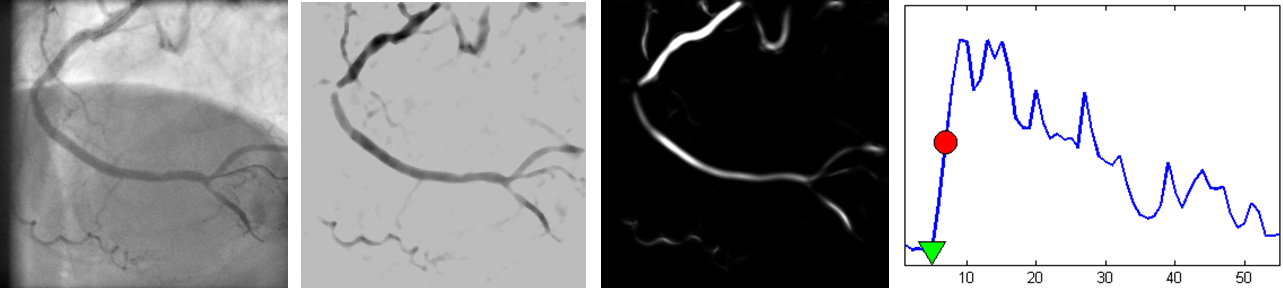}
	\caption{An example to illustrate the RNN-based method. From left to right are the original XA frame (left), the vessel layer after layer separation (middle left), the vesselness image (middle right), the contrast signal for the whole sequence (right). The color markers in the signal show the prediction of BCF with LSTM (red) and the ground truth (green). Note that the artefact of diaphragm does not appear in the vesselness image thanks to layer separation.}
	\label{fig:rnn-example}
\end{figure}

\noindent Fig. \ref{fig:rnn-example} shows an example to illustrate steps in the RNN-based method. The statistics of the absolute errors made by the three methods are shown in Table \ref{tab:compare}. The results of the mean and median errors show that the two proposed approaches have smaller errors than the state-of-the-art method, especially, the RNN-based method is able to achieve a median absolute error of 2 frames. The median of non-absolute errors (prediction minus ground truth) indicates the prediction bias of each method. The method of Condurache et al. makes late predictions, while the others have a minor bias. The table also lists the number of sequences with a small prediction error (3 frames, being about 0.2 seconds) and a large error ($>$10 frames). The method of Condurache et al. has mis-detection on 7 sequences (the first entry in the last two columns in Table \ref{tab:compare}), which was also reported in \cite{hoffmann2015robust}. While the two proposed methods both have 55 sequences with a small error ($\leqslant$ 3 frames) out of 80, the CNN-based approach has the smallest numbers of sequences with a large error ($>$ 10 frames) among the three methods.

\begin{table}[t]
	\centering
	%\begin{tabular}{l c c c c} % for diff generation, choose this option
	\begin{tabu} to \textwidth {@{} l *4{X[c]}@{}} % if a table that spans the whole page width, then choose this option
		\toprule
		Methods       &  mean (std)  &  median (*)  &  \#(error $\leqslant$ 3)  & \#(error $>$ 10) \\
		\midrule
		Condurache et al.\cite{condurache2004fast} &  6.2 (7.1)  &  5 (4)  & 29 / 73  & 10 / 73\\
		CNN-based        &   3.9 (4.9)  &  2.5 (1)  & \textbf{55} / 80 & \textbf{5} / 80 \\
		RNN-based      &   \textbf{3.6} (4.6)  &  \textbf{2} (-0.5)  & \textbf{55} / 80 & 7 / 80 \\
		\bottomrule
	%\end{tabular}
	\end{tabu}
	
	\caption{The statistics of the absolute error for the 3 methods. The two columns in the middle show the mean, standard deviation, median of the absolute errors and the median of non-absolute errors (*) in frames. The last two columns show the number of sequences on which the method made an absolute error no larger than 3 frames or larger than 10 frames.} 
	\label{tab:compare}
\end{table}

% what the results say; relate, compare to other results
\indent The median error of the RNN-based method is similar to the results reported in \cite{chen2011robust}. While they achieved a mean error of less than one frame, their detection step requires the knowledge of complete sequences, hence it will not work in a prospective scenario. The learning-based method in \cite{hoffmann2015robust} can be used for prospective detection, but some of the proposed features were heuristically designed for X-ray images of LA for EP procedure, which have different image features from the XA of coronary interventions. Compared to these methods, our approaches were designed for prospective settings and the CNN-based method is a general framework that could potentially be applied in different clinical procedures.

% unexpected result, enhancement better than learning
\indent The RNN-based learning with a handcrafted feature has slightly lower mean and median error than the CNN-based method, although the latter has a more complex and deeper architecture. This might contradict to what is commonly known about the performance of deep learning. The possible reasons may be two-fold. First, the size of training data was small, even with data augmentation and a reduced CNN model, some over-fitting was observed. Second, the CNN treats frames independently rather than modeling their temporal relations. Although CNNs perform excellent in many classification tasks, detecting BCF requires a classifier that has good accuracy for data on the border between two classes.

% Comments on computation times
\indent In terms of computation efficiency (test time), the method of Condurache et al. needed 111 ms to 443 ms to process a frame. While the CNN-based method ran very fast and used on average only 14 ms to process one frame. The RNN-based method ran on average 64 ms/frame on images of the original size $512\times512$ or $1024\times1024$, and 140 ms/frame on images of the original size $776\times776$. As the test time of the RNN-based method was based on a MATLAB implementation with a single CPU core, it has large potential to run in real-time ($<$66 ms) with an optimized implementation running on a modern GPU.

% Conclusion
%\subsubsection{Conclusion}
\indent In conclusion, we have developed two novel approaches for prospective detection of contrast inflow in XA sequences, a CNN-based and a RNN-based approach. The proposed methods perform well in BCF detection tasks in XA sequences, and outperform a previous state-of-the-art method. Both methods work in prospective settings and run fast, therefore have the potential to be integrated in advanced image guidance systems for PCI.\\

\noindent \textbf{Acknowledgement} This work was supported by Technology Foundation STW, IMAGIC project under the iMIT program (grant number 12703).\\

\noindent \textbf{Conflict of interest} The authors declare that they have no conflict of interest.

%
% -----Discussion
%
%\input{discussion.tex}
% References

%
%
%

%\input{subjidx.ind}
\end{document}